# Nanotransfer Printing of Organic and Carbon Nanotube Thin-Film Transistors on Plastic Substrates

D. R. Hines, S. Mezhenny, M. Breban and E. D. Williams[a]
*Laboratory for Physical Sciences and Department of Physics, University of Maryland, College Park, Maryland 20742.*

V. W. Ballarotto
*Laboratory for Physical Sciences, University of Maryland, College Park, Maryland 20740.*

G. Esen, A. Southard and M.S. Fuhrer
*Department of Physics and Center for Superconductivity Research, University of Maryland, College Park, Maryland 20742.*

a) Electronic mail: edw@umd.edu

A printing process for high-resolution transfer of all components for organic electronic devices on plastic substrates has been developed and demonstrated for pentacene (Pn), poly (3-hexylthiophene) and carbon nanotube (CNT) thin-film transistors (TFTs). The nanotransfer printing process allows fabrication of an entire device without exposing any component to incompatible processes and with reduced need for special chemical preparation of transfer or device substrates. Devices on plastic substrates include a Pn TFT with a saturation, field-effect mobility of 0.09 $cm^2(Vs)^{-1}$ and on/off ratio approximately $10^4$ and a CNT TFT which exhibits ambipolar behavior and no hysteresis.

Immense interest in organic thin-film electronics has lead to the demonstration of organic electronic transistors[1,2], photocells[3,4], radio-frequency circuits[5] and light-emitting diodes[6]. However, materials processing methods are still needed to assemble a broad range of high-quality organic semiconductor and metallic components onto flexible substrates[7]. Here we demonstrate that the basic concepts of nanoimprint lithography (NIL)[8] and transfer printing[9-11] (TP) can be adapted to the transfer of all the components (metallic, dielectric, and semiconductor) needed to create a high-quality, thin-film transistor (TFT) on a plastic substrate. Successful transfer printing, or lamination, requires substrates with carefully chosen differential adhesion energies[12,13] or special chemical treatment of one or more of the interfaces[14]. The heat and pressure used in NIL also provides a mechanism for facilitating transfer printing of organic materials that can be processed near or above their glass transition temperatures. We have implemented such nanotransfer printing of patterned metal, small molecule, polymer and carbon nanotube (CNT) thin films onto plastic substrates using a NX2000 imprintor (Nanonex).

In this work, transfer substrates were Si with either a native oxide surface or a 300 nm thermal oxide surface and the device substrates were plastic. Gold (Au) films, fabricated on transfer substrates using photolithography or e-beam lithography, were transfer printed onto latex, nitrile, polyvinyl chloride (PVC), polyethylene terephthalate (PET) and poly(methyl methacrylate) (PMMA) surfaces without chemical pre-treatment. As an example, 200 nm wide x 50 nm thick Au lines transfer-printed onto a PET substrate are shown in Fig. 1(a). Additionally, pentacene (Pn) films, patterned using shadow-mask deposition, were transfer printed onto photoresist (OIR908-35), latex, nitrile, PVC, PET and PMMA surfaces, again without chemical pre-treatment. In addition to sequential transfer of single-layer films, multi-layer films were also transfer printed. As an example, the transfer of a gold-on-pentacene (Au/Pn) structure fabricated by shadow mask deposition is shown in Fig. 1(b).

Physical insight into the strength of adhesion can be obtained using the Dupré equation $E_A^{AB} = \gamma_A + \gamma_B - \gamma_{AB}$ where $E_A^{AB}$ is the interfacial binding energy between materials A and B, $\gamma_{A(B)}$ is the surface free energy of A(B) and $\gamma_{AB}$ is the interfacial free energy[13,15]. Transfer of material A from substrate B to substrate C requires:

$$E_A^{AC} > E_A^{AB} \tag{1}$$

In evaluating $E_A^{AB}$ and $E_A^{AC}$ estimates based on the dispersive and polar components of the surface tension can be used. Broadly interpreted, two materials which both have either strong dispersive (oleofilic) or strong polar (hydrophilic) components are likely to adhere well[16]. For instance, a $SiO_2$ surface can be characterized as polar ($\gamma_{tot}$ = 287 mJ/m$^2$, $\gamma_D$ = 78 mJ/m$^2$, $\gamma_P$ = 209 mJ/m$^2$), whereas an organic surface can be characterized as dispersive (e.g. for PET, $\gamma_{tot}$ = 47 mJ/m$^2$, $\gamma_D$ = 33 mJ/m$^2$, $\gamma_P$ = 10 mJ/m$^2$)[13,17]. Noble metals like Au, which do not form a polar oxide, adhere weakly to $SiO_2$; however, their electrons are highly polarizable, promoting stronger adhesion to materials with dispersive surfaces such as PET. Quantitative understanding of adhesion characteristics is also likely to require consideration of mechanical linking[13] due to the heat and pressure on the plastic substrate during the transfer process.

Guided by Eq. (1), we have established a procedure (see Fig. 2) to sequentially assemble a Au gate electrode, a PMMA dielectric layer, Au source/drain (S/D) electrodes and three semiconductor films from different materials classes [Pn (small-molecule), poly (3-hexylthiophene) (P3HT) (polymer) and CNT (macromolecule)] all onto PET substrates (DuPont MELINEX® 453).

For the device fabrication, we first transfer printed (at 400 psi and 140 °C for 5 min.) a 100 nm thick Au gate electrode onto a PET device substrate from a Si substrate with 300 nm etch-back around the Au (see Fig. 2(a)). Gold S/D electrodes were then patterned on a new transfer substrate and a 200 nm thick layer of PMMA was spin coated onto both substrates. The PMMA and S/D electrodes were then transferred (at 200 psi and 140 °C for 3 min.) as shown in Fig. 2(b). Finally, the semiconductor film was patterned onto a new transfer substrate and printed (at 100 psi and 100 °C for 3 min.) as shown in Fig. 2(c). Additionally a CNT device was fabricated where the CNT film was directly printed onto the PET substrate (at 400 psi and 140 °C for 3 min.) followed by printing of the electrodes. All electrical measurements were performed at room temperature in ambient atmosphere.

A key point of the processes shown in Fig. 2 is that no lithography is performed on the device substrate, ensuring that the plastic and device components are not exposed to incompatible processes. An optical image of a typical Pn device is shown as an insert in Fig. 3. As printed (from a 50 nm Pn film thermally evaporated onto a transfer substrate), this bottom-gate, bottom-contact TFT device has not suffered the growth issues that occur with typically reported bottom-contact devices where part of the Pn layer is deposited over metal electrodes[18]. The current-voltage ($I_d$-$V_d$) curves in Fig. 3 clearly show a classic TFT response as a function of gate voltage and correspond to a saturation, field-effect mobility of 0.09 cm$^2$(Vs)$^{-1}$ and an on/off ratio approximately 10$^4$ (consistent with Pn transistors that we fabricated as bottom-gate, top-contact devices on a SiO$_2$/Si substrate).

The output characteristics of a P3HT TFT device (spin cast as a 20 nm thick film onto a transfer substrate from a 1 wt% solution on chloroform) transfer-printed onto a PET substrate are shown in Fig. 3b. Spin casting (as opposed to solution casting) and ambient conditions are both known to dramatically affect the mobility and on/off ratio in this material[19] and are likely responsible for the poor on/off ratio and transistor action observed in Fig. 3(b). Nevertheless, the variation of conductance with gate voltage for a p-type semiconducting film is clearly evident.

A bottom-gate, bottom-contact CNT TFT device fabricated via nanotransfer printing from a film deposited on a transfer substrate by chemical vapor deposition (CVD)[20] is shown in the SEM and AFM inset images of Fig. 4. In the SEM image, the CNTs show up as dark contrast due to charging of the insulating PMMA[21]. In the AFM image, the CNTs are clearly seen crossing the boundary onto the Au electrode. The gate-voltage dependence of the current at two drain voltages is shown for a top-gate, top-contact CNT device. The current increases for both positive and negative gate voltages consistent with ambipolar behavior. Previous reports of CNT films on SiO$_2$ and polyimide (with no gate electrode and transferred by etching the original SiO$_2$ support) found only p-type behavior[22,23]. The large hysteresis in gate voltage observed in CNT devices on SiO$_2$[24,25] is not observed in this device. The current in this transistor does not go to zero due to the presence of metallic CNTs in the film; high on-off ratios would require processing capable of selecting only the semiconducting tubes.

In conclusion, we have demonstrated working transistors of Pn, P3HT and CNT films assembled via sequential nanotransfer printing of electrodes, dielectric layer and semiconductor thin film onto a PET substrate. Even without process optimization, these devices demonstrate quality comparable to or better than those fabricated using standard techniques on inorganic (e.g. $SiO_2$/Si) substrates. We expect this simple, powerful technique to find broad applicability in the assembly of extensive combinations of patterned films and device substrate materials. To accomplish a predictive understanding of material transfer, the relative importance to adhesion of a materials surface free energy and mechanical modification of the interface during transfer must be determined. In addition, the possibility that the electronic properties of interfaces after transfer printing may be improved over more traditional methods of fabrication needs further study.

**Acknowledgment:** This work has been supported by the Laboratory for Physical Sciences.

**Figures & Captions:**

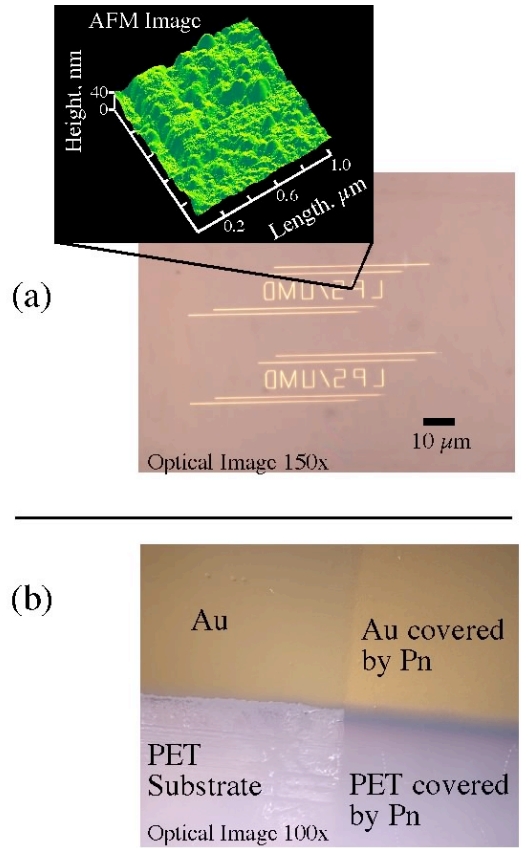

FIG. 1.

D. R. Hines hines@lps.umd.edu Applied Physics Letters

FIG. 1. (Color online). Optical images of features transfer-printed onto a PET substrate. a) 200 nm wide x 50 nm thick Au lines (insert: 1 $\mu$m x 1 $\mu$m AFM scan), b) 50 nm Pn/ 100 nm Au features transfer printed as a double layer onto a PET substrate.

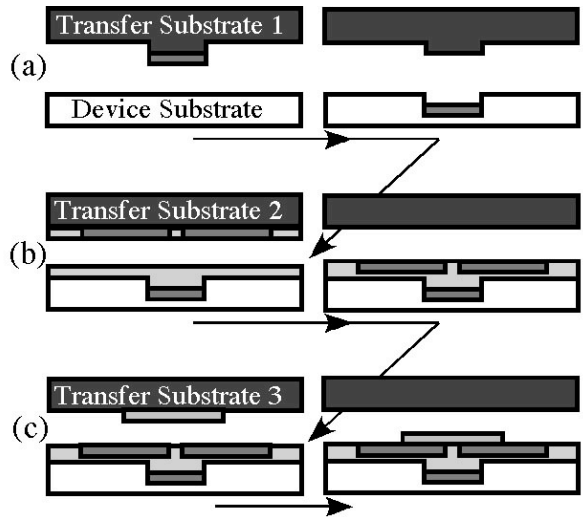

FIG. 2.

D. R. Hines              hines@lps.umd.edu             Applied Physics Letters

FIG. 2. (Color online). Transfer printing process for the fabrication of a semiconductor TFT on a PET substrate. a) printing of an embedded gate electrode, b) printing of the gate dielectric and S/D electrodes, and c) printing of the semiconductor thin film.

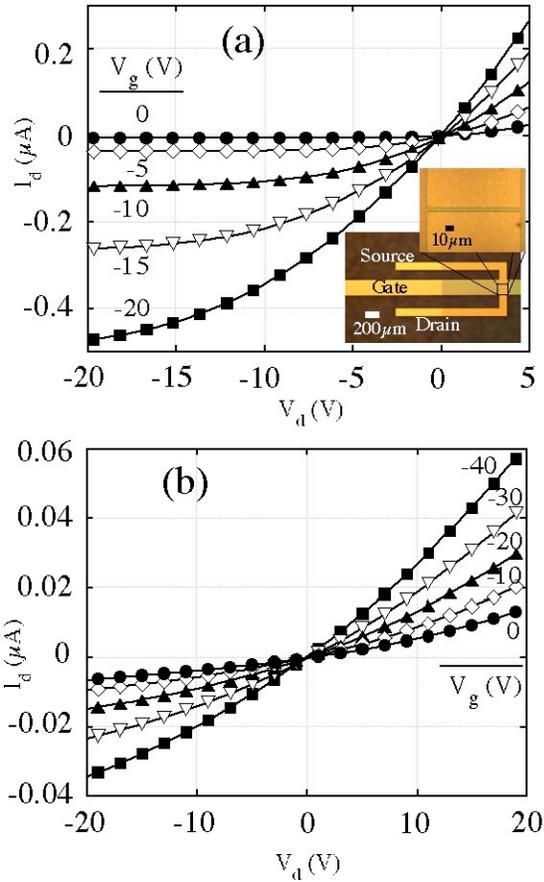

FIG. 3.

D. R. Hines            hines@lps.umd.edu            Applied Physics Letters

FIG. 3. (Color online). Current-drain voltage curves at various gate voltages for a) Pn and b) P3HT TFTs fabricated on a PET substrate via nanotransfer printing. Insert: optical image of a typical Pn device - edge of Pn layer is visible vertically across electrodes near the center of the low magnification image. Note: the device in (b) exhibited a measurable gate leakage (as large as 8 nA at Vg = -40 V); the leakage current was subtracted from each curve to obtain the data shown.

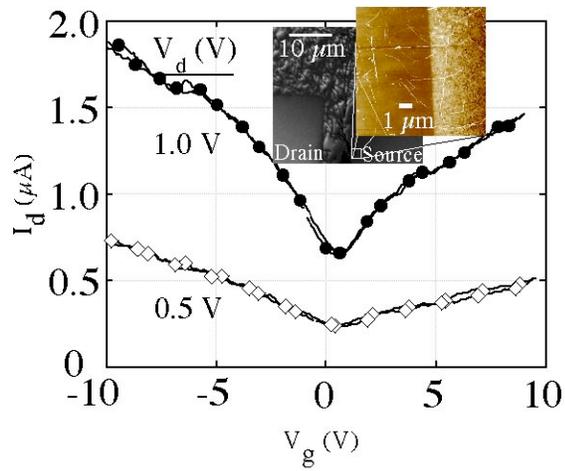

FIG. 4.

D. R. Hines              hines@lps.umd.edu              Applied Physics Letters

FIG. 4. (Color online). Current-gate voltage curves at fixed drain voltages of 1.0 V (top curve) and 0.5V (bottom curve) for CNT TFT. Inset shows SEM and AFM images of the CNT film in the source-drain region after transfer printing onto a PMMA-coated PET substrate.